CdO/MgO superlattices grown by MBE as cubic CdMgO quasi-alloys.

E. Przeździecka[1], A. Wierzbicka[1], P. Dłużewski[1], I. Sankowska[2], K. Morawiec[1], M.A. Pietrzyk[1], A. Kozanecki[1]

[1]Polish Academy of Sciences, Institute of Physics, Al. Lotników 32/46, Warsaw, Poland
[2]Łukasiewicz Research Network - Institute of Electron Technology, Al. Lotników 32/46, Warsaw, Poland

Abstract:
New perspective cubic quasi-alloys CdO/MgO short period superlattices were grown on sapphire substrates by plasma assisted molecular beam epitaxy. Their crystal quality was characterized using High Resolution X-ray Diffraction (HRXRD) and Transmission Electron Microscopy (TEM) techniques. The thickness and growth rate of MgO and CdO individual layers have been extracted. Small angle X-ray diffraction peaks corresponding to the period of the superlattices ranging from 0.6 to 5 nm were clearly observed. From the XRD measurements we obtain 65% relaxation to the substrate of superlattice (SL) in Sample A (CdO/MgO - 4.7 nm/1 nm) and 96% relaxation of SL in Sample B (CdO/MgO - 2 nm/4 nm). Surface roughness parameters for SL were obtained both by Small angle X-ray diffraction and Atomic Force Microcopy studies.

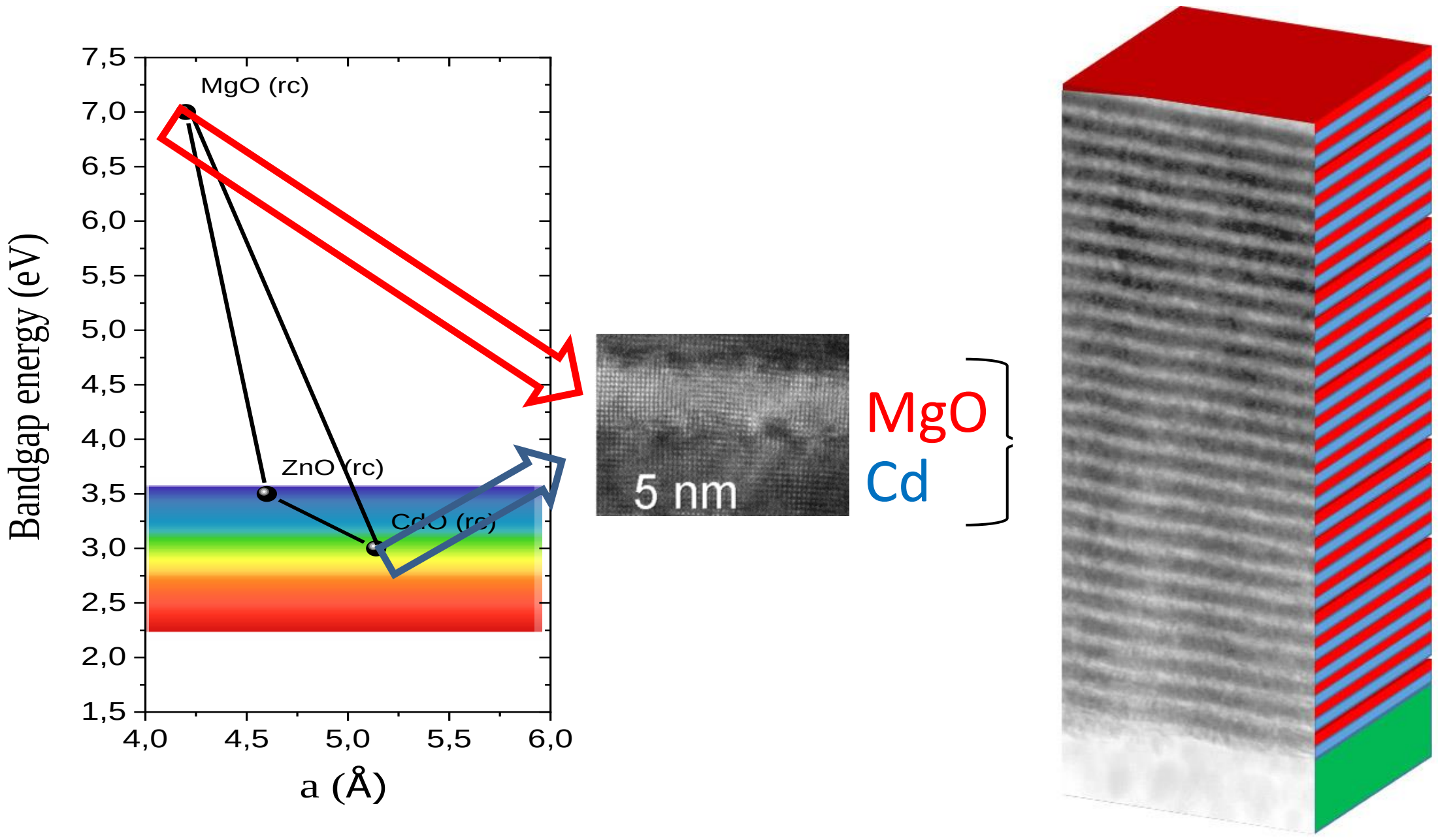

## INTRODUCTION

Pure CdO is a promising transparent conducting oxide (TCO) which exhibits metal-like charge transport behavior with an exceptionally large carrier mobility[1]. Recently, it has received a growing attention as it was shown that it can be doped with indium to achieve electron concentrations higher than $10^{21}$ $cm^{-3}$ with mobilities exceeding 150 $cm^2V^{-1}s^{-1}$.[2] The high electron mobility in CdO is a direct consequence of the very large value of the dielectric constant ($\varepsilon_o$=22) and the effective screening of ionized impurities potentials and reduced electron scattering[3]. The less-efficient electron scattering is also responsible for the reduced free-carrier absorption, making CdO a unique material with very good transparency in the infrared spectral region.

Many methods have been adopted to grow CdO films such as spray pyrolysis[4], sputtering[5],[6] pulsed laser deposition[2], MOCVD[2], sol–gel[7] etc. Cadmium-containing thin films are used in a wide variety of applications such as photovoltaic active layers[8], gas sensors[9] and others. The widespread use of TCOs as transparent electrodes in many advanced technology applications has led to intense investigations of their physical, chemical and optoelectronic nature. Importantly, improved TCO materials could have enormous impact on next-generation solar energy systems, and future progress in this field will require dedicated efforts. Good transparency in a wider spectral region of Cd based TCO can be achieved for example by doping. The transparency in the higher energy range can be increased also by preparing ternary alloys like CdMgO. Adding Mg to CdO will increase the band gap from 2.3 eV for CdO[6] to the UV range up to 7.8 eV for pure MgO. However, there are only few works studying structural, morphological, optical and electrical properties of CdMgO thin films. Recently, Gupta et al.[10] has studied the effect of growth parameters on Mg-doped CdO films grown by pulsed laser deposition (PLD). In order to obtain CdMgO mixed crystals the samples were annealed at a temperature of about 800°C for a long time ~8 h. CdMgO layers were also successfully grown by Metal Organic Chemical Vapor Deposition[11], but in a wide composition range only CdMgO nanoparticles were received by spray pyrolysis technique[12]. Theoretically, the rocksalt cubic structure is stable over all (Mg,Cd)O compositions[13]. However, in the case of the growth of layers, the whole range of concentrations without phase separation has not been obtained yet. Besides this, studies on (CdO/MgO) superlattices (SL) and MgCdO films grown by MBE technique are scarce in the literature. Moreover, layers with high concentrations of Mg in CdMgO have been less studied in the current literature probably due to intrinsic difficulties in making the different growth conditions for the two binary CdO and MgO compounds compatible. Usually, CdO is grown in much lower temperatures than MgO. Short period SLs can solve the problem of phase separation and non-homogeneity of the ternary material. In the TCO family II−VI oxides ternary alloys have attracted considerable interest of the scientific community due to the possibility of modulating their interesting optoelectronic properties.

Here, we explore short period CdO/MgO superlattices (SLs) which were grown by plasma assisted molecular beam epitaxy (PA-MBE) on r-oriented sapphire substrates. There are no reports in the literature about this kind of structures. It is a novel type of superlattices which can be used to obtain high quality TCO layers.

## EXPERIMENTAL

Two types of CdO/MgO SLs were grown by PA MBE in a RIBER Compact 21 machine on r-oriented sapphire substrates. The SL structures consist of 30 pairs – sample "B" and 25 pairs – sample “A” of CdO/MgO bilayers (Fig. 1). The growth parameters (fluxes and temperatures) for both types of SLs were the same. The structures were deposited directly on sapphire without any buffer layers.

Prior to the growth, the sapphire substrates were chemically cleaned and then annealed at a temperature of about 700°C in high vacuum for 30 min and finally in oxygen plasma also in 700°C. The growth temperature was about 360°C. The oxygen plasma parameters were as follows: power 450 W and 3 ml./min oxygen flow.

Topography of samples was analyzed with a Vecco Icon Atomic Force Microscopy system. TEM experiments were carried out using a Titan Cubed 80-300 microscope operating at 300 kV. Focused Ion Beam (FIB) technique was used to prepare cross-sectional samples.

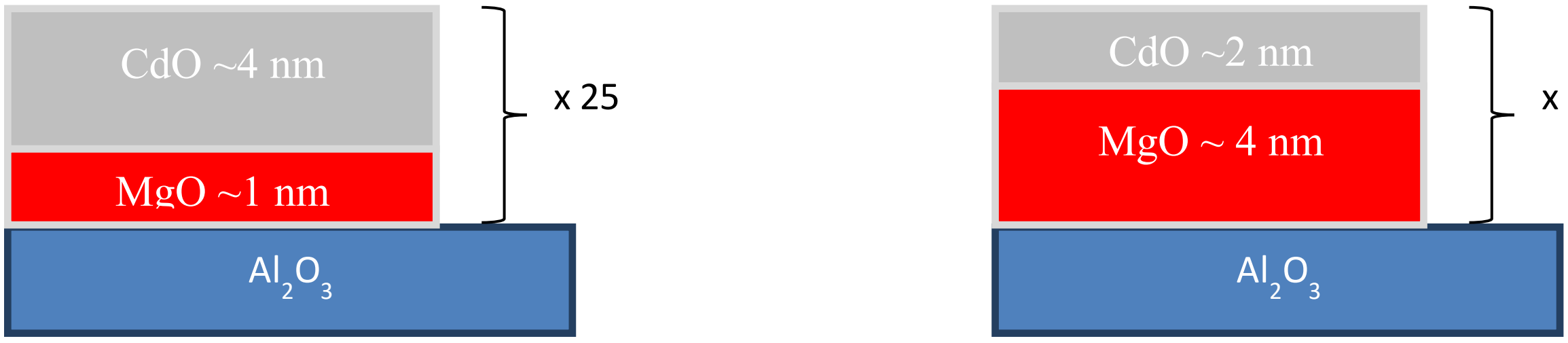

Fig. 1 Schematic figures of CdO/MgO superlattices (a) sample “A” (b) sample “B.

The X-ray diffraction (XRD) measurements were performed using a Panalytical X'Pert Pro MRD diffractometer equipped with a hybrid two-bounce Ge (220) monochromator, triple-bounce Ge (220) analyzer (or Soller slits) in front of the Pixcel detector. First θ/2θ scans in low angle resolution were executed in a wide range of 2θ angle. Next high resolution XRD measurements (2θ/ω scans and high resolution XRD reciprocal space maps (RSMs)) were performed using analyzer in front of detector.

## RESULTS AND DISCUSSION

Although the samples were grown at a relatively low temperature (360°C on the thermocouple) their good crystal quality of SLs structures is achieved as demonstrated by XRD. Fig. 2. (a, b) shows the full-range XRD patterns of CdO/MgO SLs. The θ/2θ scans confirm the rhombohedral type structure of the substrate oriented in [01-12] direction (r-orientation) and cubic structure of the CdO/MgO superlattices. As shown in Figure 2 (a, b), the 200 diffraction peaks come from the CdO/MgO SLs, and high orders of superlattice-related satellite peaks are clearly observed, confirming the good periodicity and smoothness of the interfaces. Zero order peaks describing average parameters of SLs are marked as $S_0$ (Fig. 2 c, d ). Periods of these structures can be calculated from higher order peaks. For sample A the SL period is equal to 5.2 ± 0.1 nm, and for sample B 6.0 ± 1.0 nm. The satellite peaks are well defined in both samples. When the thickness of MgO layers increase in SL (sample “B”), the interfacial quality of superlattice is deteriorated remarkably, as evidenced in Fig. 2b. The order of the satellite peaks is lower and the intensities of the satellite peaks are much weaker compared to the sample “A”. Fig. 2c, and d show the high angle resolution XRD 2θ-ω scans of both samples. Sample “B” with a CdO/MgO layer have broader SL-related peaks, while sharp satellite peaks are observed in sample “A” with thicker CdO layers and thinner MgO layers. In order to determine the precise value of $a_{\perp}$, and to get more information about the quality and relaxation relation in SLs we measured symmetrical 200 CdO/MgO reciprocal space maps (RSM) (Fig. 3 a, c) for both structures. Also lattice parameters were measured in the perpendicular direction to check tetragonal distortion of the elementary cells. The value of $a_{\|}$ was determined from asymmetrical RSM of 412 reflection of CdO/MgO SL (Fig.3 b, d). Based on the presented data the perpendicular and parallel lattices parameters were calculated and they are $a_{\|}$ = 4.6588 Å; $a_{\perp}$ = 4.7034 Å for sample B and $a_{\|}$ = 4.8069 Å, $a_{\perp}$ = 4.6731 Å for sample A.

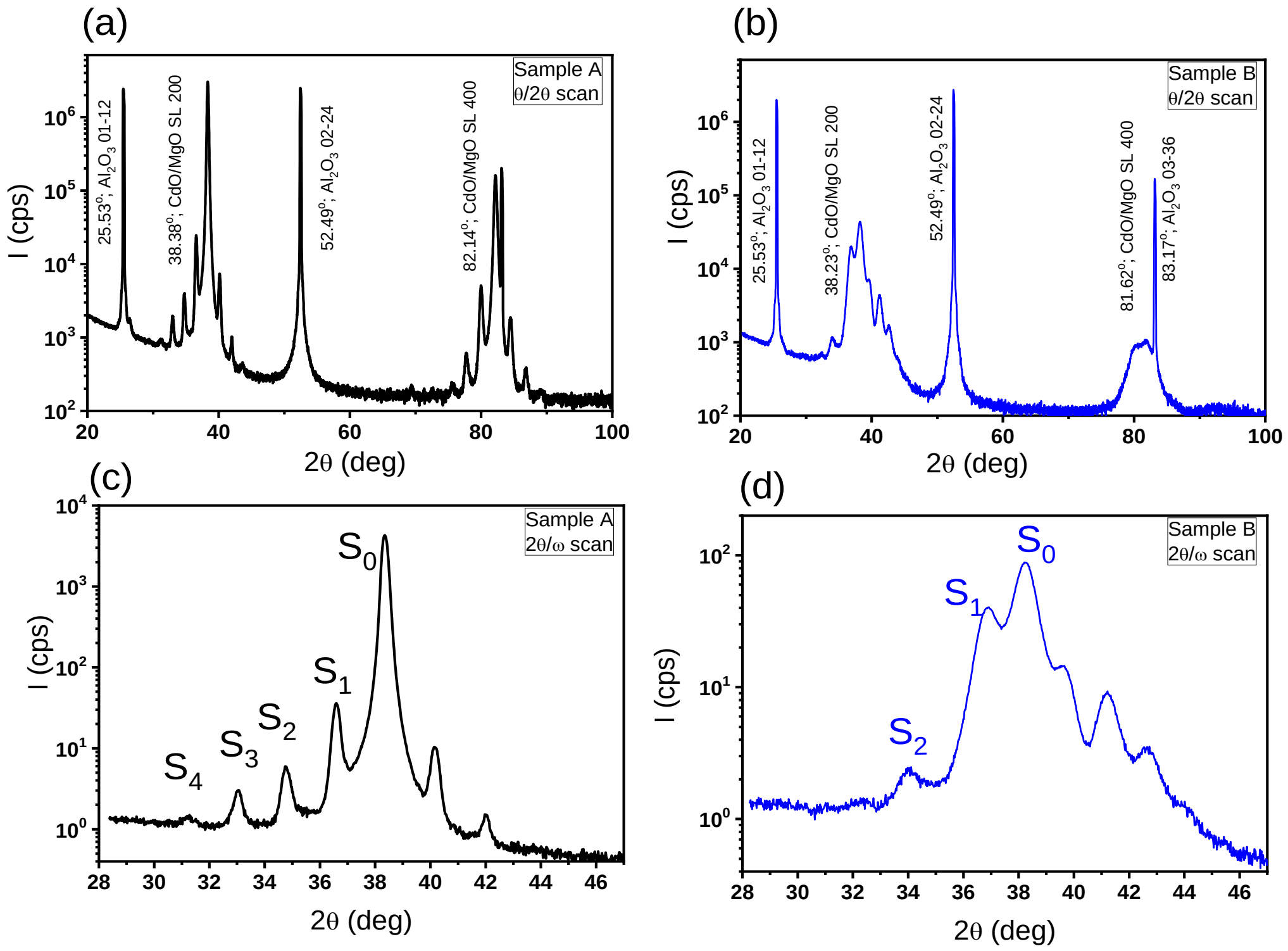

Fig.2. 2Theta/Omega high angle resolution XRD scans (c, d) and low angle resolution Theta–2Theta (a, b) XRD patterns of the CdO/MgO superlattices.

The symmetrical 200 CdO/MgO RSMs show peaks with broadening in the $Q_x$ direction which indicates lower structural quality and high dislocation density of sample “B”. Broadening of the zero order peak coming from the SL indicate high diffuse scattering signal with is related to high dislocation density[14]. Much better structural quality is observed in case of sample “A” (Fig. 3 (a, b)). Asymmetrical 412 CdO/MgO RSMs shows that SLs are partially relaxed to sapphire substrate. We calculated that relaxation to the substrate is 65% for sample A and 96% for sample B. It is the evidence of better structural quality of Sample A.

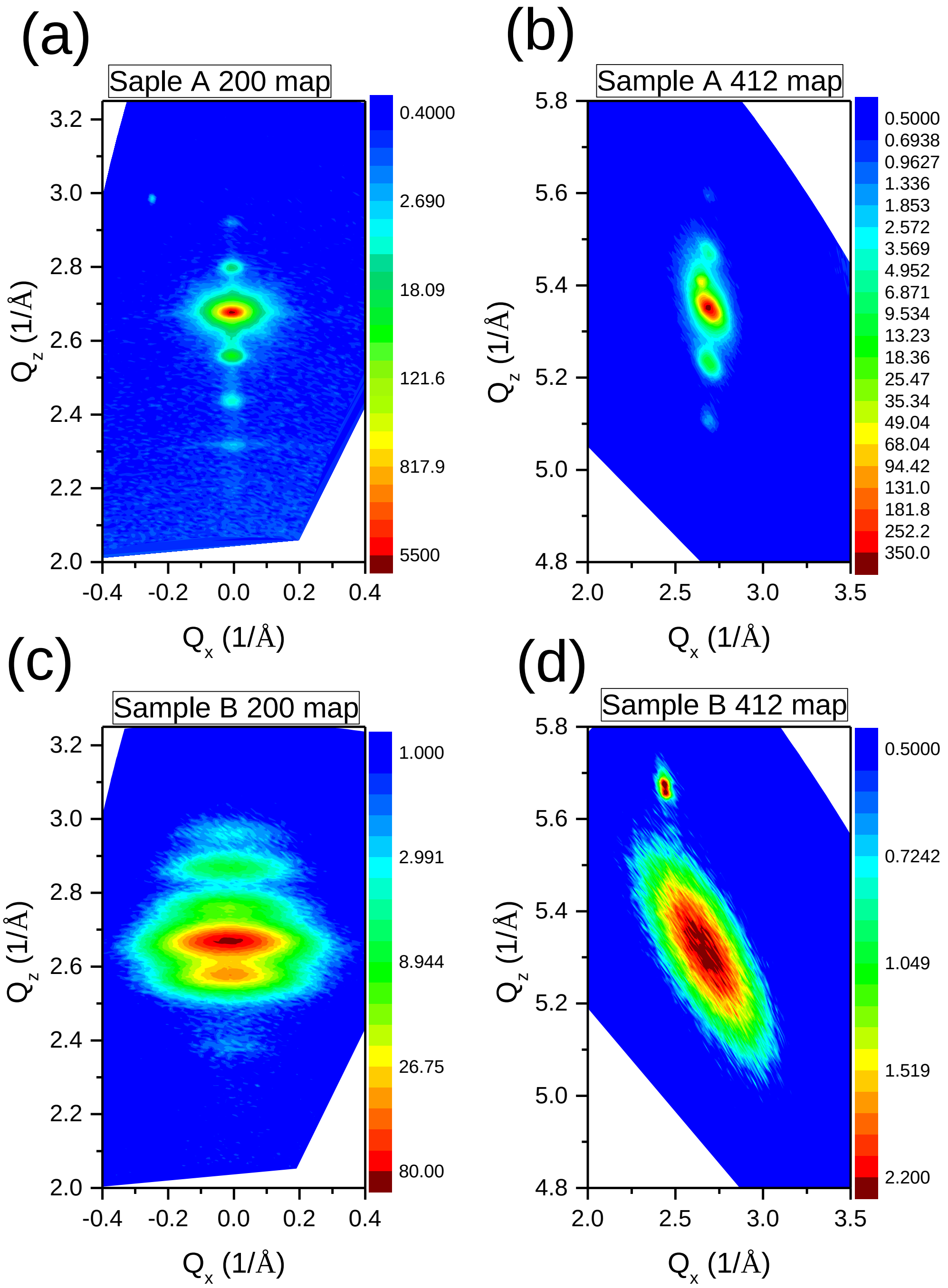

Fig. 3. Reciprocal space maps from the CdO/MgO superlattice sample around the 200 (a and c) and 412 Bragg peaks (b and d).

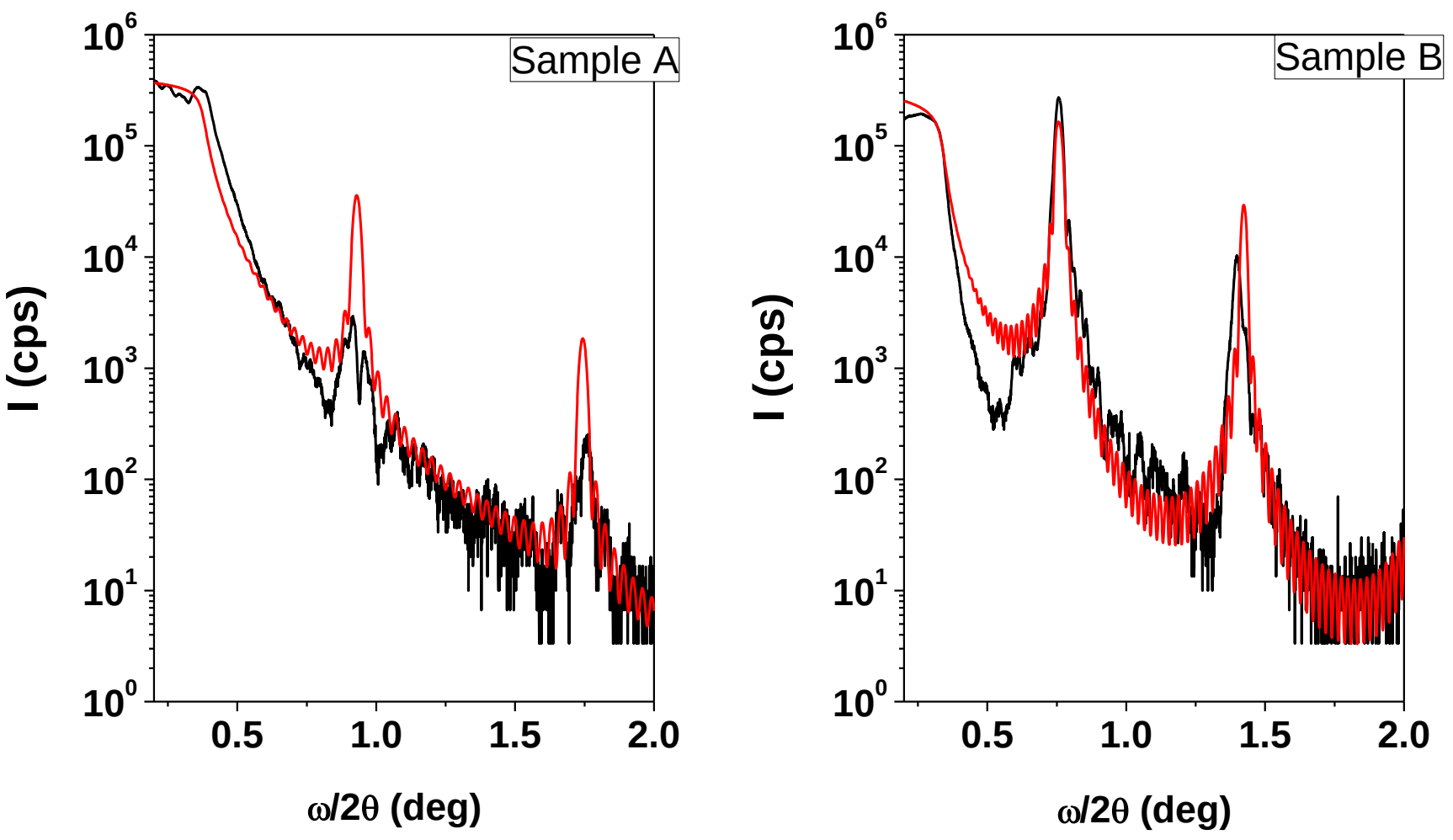

Fig. 4. Experimental (black line) and simulated (red line) low angle x-ray reflectivity curves from as-deposited CdO/MgO superlattices.

Individual layer thicknesses, $D_{MgO}$, and $D_{CdO}$, in SLs were determined by simulations of the x-ray reflectivity (XRR) curves (Fig. 4). The simulated data are presented in Table 1. The reflectivity curves are shown in Fig. 4 together with the simulation results. Oscillations on the reflectivity curves due to the CdO/MgO interfaces, commonly known as Kiessig fringes[15], are clearly visible on the XRR profiles. Their presence shows the high quality of the interfaces and allows to determine the thickness of the layers. Film density can be determined from the angle at which X-rays begin to penetrate the sample - the so called critical angle for external reflection. Below some critical incident angle ($\theta_c$), dictated by Snell's law, the incident beam will be totally reflected from the surface. Below the $\theta_c$ the X-rays begin to penetrate the sample and the collected intensity decreases. The critical angle position is related to the bulk density of the material. Experimental curves were simulated using commercial Panalytical software, based on Parrat formalism for reflectivity. The approach used is detailed in refs[16–18]. Densities determined by computer simulations are: 9.1 g/cm$^3$ for CdO layers and 3 g/cm$^3$ for MgO layers, consistent with the analysis of the critical angle position.

Root-mean-square (RMS) roughness parameters can be also extracted from XRR curves. In order to test the validity of the method demonstrated above, the obtained lattice parameters, layer thicknesses and interface width are compared to AFM and TEM results. Surface roughness parameters for SLs are presented in Table 2 and compared to the values extracted from AFM measurements. Despite the fact that RMS for AFM comes from the area 2×2 μm and from XRR from the area larger than 1.4 mm (irradiated length) × 0.5 mm the obtained results are comparable. Thickness parameters extracted from reflectivity curves by computer simulation are reported in Table 1 and the values of the layers thickness are compared to those obtained from TEM.

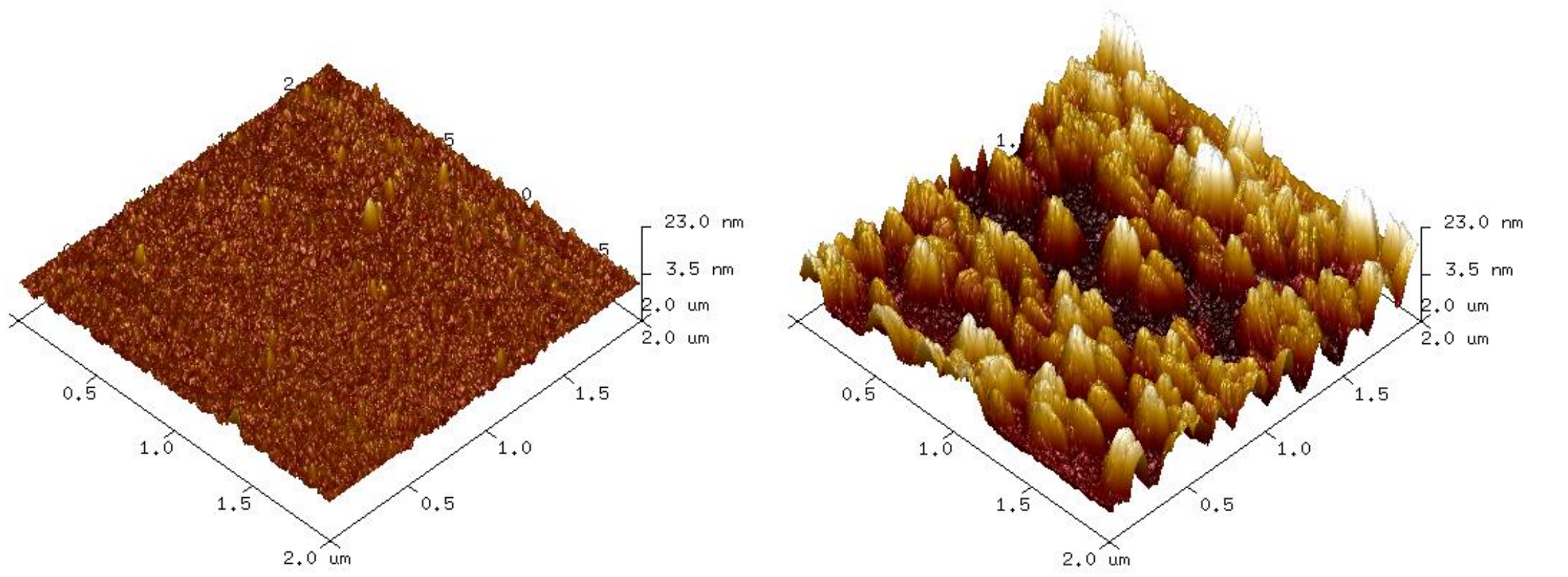

Fig. 7. AFM images of sample A and B.

| Samples | Time (min) CdO/MgO | XRR(nm) CdO/MgO | TEM (nm) CdO/MgO | Growth rate from XRR CdO/MgO (nm/min) | Growth rate From TEM CdO/MgO (nm/min) |
|---|---|---|---|---|---|
| B | 4/4 | 1.87/4.53 | 2/4 | 0.46/1.13 | 0.5/1 |
| A | 6/1 | 4.2/0.99 | 4.7/1 | 0.7/0.99 | 0.78/1 |

Table. 1. Layers thickness and growth rates based on XRR and TEM measurements.

| Samples | RMS from XRR (nm) | RMS from AFM (2x2μm) (nm) |
|---|---|---|
| B | 4.2 | 4.7 |
| A | 0.43 | 0.99 |

Table. 2. Root-mean-square surface roughness parameters extracted from XRR and AFM.

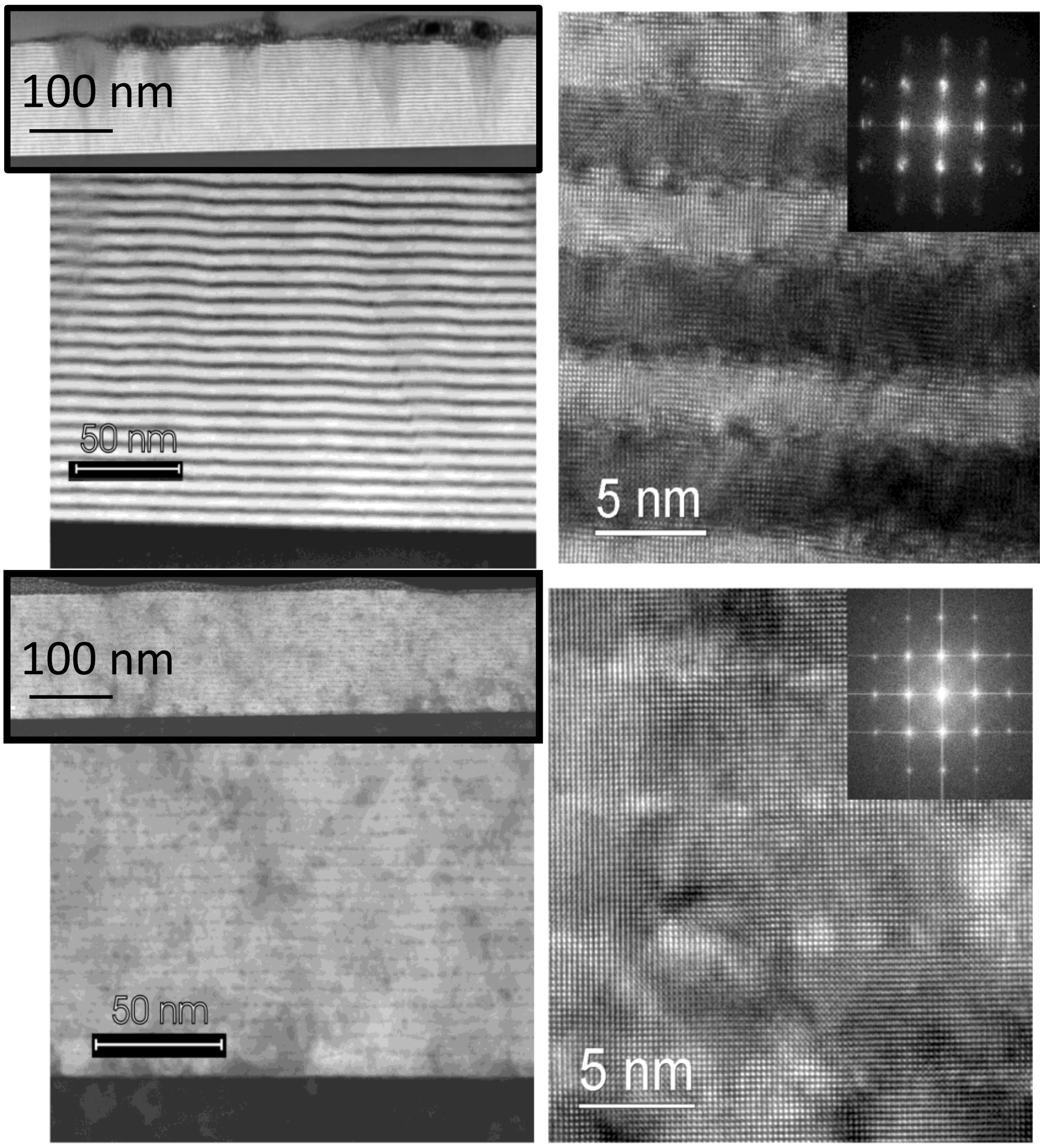

Fig. 5. STEM/HAADF (left column) and HRTEM (right column) cross-sectional images of B (upper row) and A (lower row) taken with the electron beam along the <001> zone axis. Insets show Fourier transform of HRTEM image.

TEM images of both samples confirm the presence of SL and reveal cubic structure of the layers (Selected Area Diffraction (SAED) pattern in Fig. 5). STEM / HAADF images of sample "B" revealed uniform and continuous CdO and MgO layers. It is worth to mention that SL had lateral disturbance in the form of "V" shape defects (upper left panel in Fig.5.) with an origin near the substrate and propagated up to the surface layer on which few nm jump(steps) are formed. This disorder influences XRD results and can explain widening of the peaks in this sample. HRTEM images show interface roughness at the level of 0.5 nm. Both layers consist of nano-crystalline blocks of 3-4 nm in size and misoriented of about 5°. The Fourier transform of the HRTEM images presented two rectangular patterns of spots differing in scale by about 11±2%. These patterns well matching CdO and MgO structures projected along the <100> zone axis.

In the case of sample "A" the CdO layers were of about 4.7 nm thick, whereas MgO layers were about 1 nm. The MgO layers were non-uniform and non-perfectly continuous. In this situation the interface roughness would be determined as a vertical size of the MgO thickness inhomogeneity i.e. about 1 nm. The “V” shape defects were not observed in this structure. HRTEM images showed good crystallinity of the SL without appearance of nano-blocks observed in the case of sample “B” (upper part of Fig. 5). The Fourier transform HRTEM presented only one set of rectangular spots confirming high quality of the SL crystal structure.

Based on SAED in sample “B” with higher thickness of MgO sublayers two individual sublattices are visible. Visible patterns coming from CdO and MgO layers and obtained interplanar distances correspond to the values reported for bulk MgO (2.11 Å) and bulk CdO (2.35 Å) crystals. All measured interplanar distances obtained from SAED and XRD measurements are collected in Table 3. It should be mentioned, that in case of XRD analysis the averaged interplanar distances for SL were calculated, whereas in case of SAED its depends on relaxation between two kind of layers.

Table. 3. Interplanar distances in SLs based on XRD and TEM analysis.

<table>
<tr><td colspan="4">Interplanar distances</td></tr>
<tr><td></td><td>Sample B</td><td>Sample A</td><td>Bulk</td></tr>
<tr><td>CdO</td><td>$d_{\perp}$ 2.37±0.04Å<br>$d_{II}$ 2.18±0.04 Å</td><td rowspan="2">$d_{\perp}$ 2.34±0.04 Å<br>$d_{II}$ 2.36±0.04 Å</td><td>$d_{\perp}$ 2.349 Å</td></tr>
<tr><td>MgO</td><td>$d_{\perp}$ 2.12±0.04 Å<br>$d_{II}$ 1.98±0.04 Å</td><td>$d_{\perp}$ 2.11 Å</td></tr>
<tr><td>CdO/MgO SL (XRD)</td><td>$d_{\perp}$ 2.35±0.05 Å</td><td>$d_{\perp}$ 2.34±0.05 Å</td><td></td></tr>
</table>

## CONCLUSIONS

Cubic 100 oriented CdO/MgO superlattices were successfully obtained by PA-MBE on r-sapphire substrates. X-ray reflectivity was used to provide accurate characterization of CdO/MgO multilayer films. Structural parameters determined by computer simulation of the reflectivity data were used to measure the rate of deposition for CdO and MgO at applied growth temperature and to quantify observations regarding layer density and surface roughness. Deposition rates are at about 1nm per minute in case of MgO layers and at about 0.5-0.7 nm/min. in case of CdO. The obtained surface roughnesses are comparable with the data provided by AFM. Based on TEM selected area diffraction patterns analysis almost fully relaxed superlattices were detected in the case of 2/4 nm (CdO/MgO) structures, whereas partially relaxed in case of 4.7/1 nm structures. Good-quality MgO/CdO heterostructures, with controlled MgO and CdO thickness, have been first time successfully obtained and characterized what can opening the door for their future applications.